\documentclass[sigconf]{acmart}

\acmBooktitle{Companion Proceedings of the 34th ACM Symposium on the Foundations of Software Engineering (FSE '26), June 5--9, 2026, Montreal, Canada}

\AtBeginDocument{%
  }

\copyrightyear{2026}
\acmYear{2026}
\setcopyright{cc}
\setcctype{by-nc-nd}
\acmConference[FSE Companion '26]{34th ACM Joint European Software Engineering Conference and Symposium on the Foundations of Software Engineering}{July 05--09, 2026}{Montreal, QC, Canada}
\acmBooktitle{34th ACM Joint European Software Engineering Conference and Symposium on the Foundations of Software Engineering (FSE Companion '26), July 05--09, 2026, Montreal, QC, Canada}
\acmDOI{10.1145/3803437.3804877}
\acmISBN{979-8-4007-2636-1/2026/07}

\begin{document}

\title{Reliable and Developer-Aligned Evaluation of Agents \\ for Software Engineering}

\author{Razvan Mihai Popescu}
\email{r.m.popescu@tudelft.nl}
\orcid{0009-0003-6251-770X}
\affiliation{%
  \institution{Delft University of Technology}
  \city{Delft}
  \country{The Netherlands}
}


\begin{abstract}

Large language models are rapidly moving towards closing the development cycle, transitioning from simple assistive companions to autonomous contributors deeply embedded into collaborative development environments. Despite their accelerated adoption, existing evaluation techniques are limited due to their fragmented nature and distorted projection of true model capabilities, often obtained from hypothetical syntactic scenarios. This research aims to bridge this gap by providing a comprehensive evaluation methodology for LLM-powered agents that is grounded in real-world software development practice. Our  evaluation approach focuses on contamination-awareness, in-the-wild agentic behavior assessment, and trajectory-aware benchmarks and metrics capturing realistic coding contexts, human-aligned behavior, and model failure modes.

\end{abstract}






\maketitle

\section{Introduction}

Given their strong reasoning and abstraction capabilities, Large Language Models (LLMs) have been widely employed in the context of Software Engineering (SE), which has become a prominent application domain. These models have shown a remarkable capacity to understand the structured nature of source code and achieve strong performance in various code-related tasks, leading to an increasing reliance on empirical evaluations to assess their abilities and limitations. 

However, current evaluation practices are often unreliable, difficult to reproduce, and poorly aligned with real-world developer needs. Many works strongly rely on convenience-based metrics borrowed from NLP, such as BLEU, METEOR, or ROUGE, which fail to capture functional validity and developer utility~\cite{egor2023_outofblue, roy2025_evalsumm}. At the same time, many of these metrics are inconsistently used across different coding tasks with minimal to no adaptation, while their coding variants, such as CodeBLEU, or embedding-based versions, such as CodeBERTScore, have shown to perform on par with general translation metrics~\cite{egor2023_outofblue, roy2025_evalsumm}. Furthermore, these metric-level limitations are amplified by flaws in widely used benchmarks. Various benchmarks suffer from issues such as saturation, data contamination, incorrect ground truths, and unrealistic context scenarios, which not only distort the reported performance, but also hinder the comparability and reproducibility of the studies~\cite{hu2025assessingadvancingbenchmarksevaluating}. Lastly, although LLMs are increasingly being used as evaluation judges, their assessments are prone to biases and hallucinations, often diverging from human judgements and further compromising the validity of~\mbox{evaluation protocols~\cite{Wang_2025}.} 

Nevertheless, these evaluation practices are still the norm in the community, despite often creating an illusion of model competence and resulting in misleading conclusions and potentially harmful downstream effects. The recent emergence of coding agents adds a new dimension to this evaluation landscape, as their autonomous and tool-mediated interactions with software systems challenge the assumptions of traditional one-shot evaluation setups. Therefore, there is a growing disparity between how these models are evaluated in research settings and how they are actually used by developers in practice.  
% contamination-aware dataset construction,

This research aims to bridge this gap by developing a principled, multi-dimensional evaluation approach for LLM-powered agents in software engineering. Through systematic analysis of existing evaluation practices, empirical studies of agent behavior in real-world repositories, and the design of contamination-aware benchmarks targeting software evolution and maintainability, this work seeks to contribute towards a community-facing evaluation standard that enables reliable, reproducible, and developer-relevant assessment of LLM-based systems.

% The recent emergence of autonomous coding agents adds a new dimension to the evaluation landscape, as their behavior unfolds over time and within evolving software systems, making traditional one-shot evaluation setups insufficient.

% As a result, there is a growing gap between how LLMs and agents are evaluated in research settings and how they are actually used by developers in practice.

 %  dupa zici ca oferta informatii false practic. Dupa tranzitionezi catre solutia noastra. 

% With prominent models such as LLama 3/4 or gemini still being evaluated on benchmarks such as HumanEval, despite known issues related to it. 

% The rapid adoption of Large Language Models (LLMs) in software engineering has led to an increasing reliance on empirical evaluations to assess their capabilities and limitations. However, current evaluation practices are often unreliable, difficult to reproduce, and poorly aligned with real-world developer needs. Many studies rely on static benchmarks, surface-level automatic metrics, and datasets with unclear provenance or potential training contamination, resulting in performance estimates that are fragile, misleading, or disconnected from practical software development outcomes.

\section{Related Work}

% While LLMs have been applied throughout the software development cycle, the question of how to fairly and meaningfully assess them remains open. 
Popular benchmarks such as HumanEval, MBPP, and Defects4J provide reproducible measures of functional correctness, yet are known to suffer from saturation, data contamination, and limited contextual realism issues (e.g., method-level operation) increasingly evident as frontier models, including GPT-o1 or Qwen-Coder~\footnote{https://evalplus.github.io/leaderboard.html}, report near-ceiling performance on these tasks~\cite{hu2025assessingadvancingbenchmarksevaluating, riddell2024quantifyingcontaminationevaluatingcode}. To move beyond surface-level metrics, some studies employ LLMs as judges to score or rank generated code~\cite{Wang_2025}, while others rely on controlled user studies measuring productivity or developer satisfaction, complementing these benchmarks~\cite{becker2025measuringimpactearly2025ai}. However, LLM-based judging raises concerns around bias and reproducibility, while human evaluations not only remain costly and difficult to scale but suffer from low external validity, failing to capture the complexity and pace of real-world development environments.  

% rely on human evaluations to assess qualities such as maintainability or intent alignment. However, LLM-based judging raises concerns around bias and reproducibility, and human evaluations remain costly and difficult to scale. Controlled studies measuring productivity or developer satisfaction complement these benchmarks but are limited in scale and external validity

% To assess AI assistants based on output quality, usability, user satisfaction, or productivity, prior studies rely on controlled user experiments that evaluate the relative performance of these systems in terms of the developer experience and task goals. Although this evaluation setting is strongly related to the operational nature of these tools, it suffers from low external validity, not entirely capturing the complexity, pace, and nuances of real-world development.

On the other hand, coding agents not only add a layer of autonomy and persistence over their precedents, but they are also capable of adjusting their behavior in response to observed effects while maintaining progress towards a user goal. Such agents are now able to perform multi-step task planning, take actions, run self-evaluations, and coordinate over different tools~\cite{liu2025largelanguagemodelbasedagents}. Current agentic benchmarks such as SWE-Bench, SWT-Bench, or AgentBench evaluate these agents under controlled conditions in different contexts, including issue resolution, bug reproduction, and decision making. However, these practices only capture narrow aspects of agent behavior, neglecting real-world collaboration and the ever-changing nature of development environments.

\section{Research Plan}
% In the following, we present three main steps toward achieving robust LLM evaluations in SE, including a systematic review of existing evaluation techniques, in-the-wild assessment of agentic activity, and model benchmarking within real-world evolving environments.
We outline three steps towards robust LLM evaluation in SE: a systematic review, in-the-wild assessment of agentic behavior, and benchmarking in real-world development workflows.

\subsection{Evaluation Landscape}
We first establish the empirical and methodological foundation for rethinking how LLMs should be evaluated in SE. We conducted a systematic literature review by synthesizing 279 peer-reviewed papers on 26 coding tasks in order to provide a comprehensive overview of the LLM4Code evaluation landscape, along with its limitations. We proposed a taxonomy of evaluation setups, audited the datasets and benchmarks currently in use, and tracked model adoption across the literature. By exposing the disconnect between performance scores, functional validity, and contextual clues, we establish a roadmap for rigorous, human-aligned, and functionally-sound evaluation frameworks. Our main research questions are as follows.

\begin{itemize}
    \item[\textbf{RQ1:}] Which LLMs have been evaluated in the context of code-related tasks?
    \item[\textbf{RQ2:}] What datasets are driving the evaluation of LLMs across code-related tasks?
    \item[\textbf{RQ3:}] What evaluation techniques have been applied to assess LLMs in code-related tasks
\end{itemize}

\subsection{In-the-Wild Evaluation}

The emergence of specialized coding agents powered by LLMs, including OpenAI Codex, GitHub Copilot, Claude Code, or Google Jules, opens up a new evaluation paradigm due to their tight operation within development environments and manipulation of version control systems. By exploiting their characteristic tell-tale signatures, we evaluate coding agents in open-source collaborative repositories, moving beyond the synthetic environments of conventional evaluation pipelines towards "in-the-wild" evaluations reflecting real coding workflows. We perform a thorough analysis of both agentic and human development activities, along with a longitudinal investigation of their impact on codebase maintainability together with their failure points, and release a large-scale dataset of agentic and human contributions to support further research~\cite{mihai2026investigating}.

\begin{itemize}
    \item[\textbf{RQ1:}] What is the difference between agent-authored and human-authored activity in shaping collaboration and development progress?
    \item[\textbf{RQ2:}] How do agent-authored contributions influence the trajectory of code maintenance over time compared to human-authored ones?
\end{itemize}

\subsection{Change-Responsive Evaluation}
In practice, software development typically unfolds through evolving requirements, iterative changes, and collaborative decision-making, shaping the trajectory of codebases over time. Based on these observations, this study introduces a multilingual, contamination-aware, benchmark for end-to-end issue resolution in evolving software repositories. 
Our main research questions are as follows.
% This benchmark not only evaluates the ability of models to generate correct patches, but also multi-step resolution trajectories and developer intent alignment, such as patterns observed in pull requests or code review decisions, reflecting real-world coding practices.  

\begin{itemize}
    \item[\textbf{RQ1:}] How do agents’ performance patterns evolve when iteratively responding to feedback in complex development tasks?
    \item[\textbf{RQ2:}] Are there systematic differences in behavior or failure modes across distinct parts of a software system under repeated modifications?
    % \item[\textbf{RQ3:}] How does prior resolution experience influence agent performance on analogous tasks within the same development context?
\end{itemize}

\section{Conclusion}
We propose a methodology for reliably evaluating LLM-powered agents in real-world development, focusing on human-aligned behavior, evolving environments, and failure resolution. We first systematize existing evaluation practices to identify limitations, then study in-the-wild agent behavior to uncover realistic challenges and failure modes, and finally use these insights to design benchmarks and metrics that better reflect developer intent and practice.
% This research develops a methodology for reliably assessing LLM-powered agents in real-world development settings, focusing on human-aligned behavior, environment evolution, and failure resolution.
% To this end, we first analyze and systematize existing evaluation practices to expose their limitations, then study agent behavior in the wild to surface realistic challenges and failure modes, and finally use these insights to inform the design of benchmarks and metrics that better reflect developer intent and realistic coding environments.

\bibliographystyle{ACM-Reference-Format}
\bibliography{sample-base}

% \appendix

\end{document}